\documentclass[a4paper,11pt]{article}
\usepackage[dvips]{graphicx,psfrag}
\usepackage{amssymb}

\makeatletter

 \@addtoreset{equation}{section}
\makeatother
\newcommand{\beq}{\begin{eqnarray}}
\newcommand{\eeq}{\end{eqnarray}}

\def\bra#1{\left\langle #1\right\vert}
\def\ket#1{\left\vert #1\right\rangle}

\begin{document}
%%%%%%%%%%%%%%%%%%%%%%%%%%%%%%%%%%%%%%%%%%%%%%%%%%%%%%%%%%%%%%%%%%%

%%%%%%%%% title %%%%%%%%%%%%%
\begin{titlepage}

\begin{flushright}
OU-HET 436\\
{\tt hep-th/0304047}\\
April 2003
\end{flushright}
\bigskip

\begin{center}
{\LARGE\bf
Unitarity Bound of the Wave Function Renormalization Constant 
}
\vspace{1cm}

\setcounter{footnote}{0}
\bigskip

\bigskip
{\renewcommand{\thefootnote}{\fnsymbol{footnote}}
{\large\bf Kiyoshi Higashijima\footnote{
     E-mail: {\tt higashij@phys.sci.osaka-u.ac.jp}} and
 Etsuko Itou\footnote{
     E-mail: {\tt itou@het.phys.sci.osaka-u.ac.jp}}
}}

\vspace{4mm}

{\sl
Department of Physics,
Graduate School of Science, Osaka University,\\ 
Toyonaka, Osaka 560-0043, Japan \\
}

\end{center}
\bigskip

%%%%%%%%% abstract %%%%%%%%
\begin{abstract}
The wave function renormalization constant $Z$, the probability of finding a bare particle in the physical particle, usually satisfies the unitarity bound $0 \leq Z \leq 1$ in field theories without negative metric states. This unitarity bound implies the positivity of the anomalous dimension of the field in the one-loop approximation. In nonlinear sigma models, however, this bound is apparently broken because of the field dependence of the canonical momentum. The contribution of the bubble diagrams to the anomalous dimension can be negative, while the contributions from multi-particle states satisfy the positivity of the anomalous dimension, as expected. 
 We derive the correct unitarity bound of the wave function renormalization constant.

\end{abstract}

\end{titlepage}

\pagestyle{plain}

%%%%%%    TEXT START    %%%%%%
%%%%%%%%%%%%%%%%%%%%%%%%%%%%%%%%%%%%%%%%%%%%%%%%%%
\section{Introduction}
%%%%%%%%%%%%%%%%%%%%%%%%%%%%%%%%%%%%%%%%%%%%%%%%%%

The wave function renormalization constant $Z$ satisfies the unitarity bound $0 \leq Z \leq 1$, because it is interpreted as the probability of finding a bare particle in the physical one-particle state \cite{UK, KL, Nishijima, PS, Weinberg}. In perturbation theory, this bound implies the positivity of the coefficient $c$ in the one-loop contribution to the wave function renormalization constant,
\begin{equation}
Z=1-cg^2\log{\frac{\Lambda^2}{\mu^2}}\leq 1, \qquad (c>0)\label{eqn:oneloop}
\end{equation}
where $g, \Lambda$ and $\mu$ denote the coupling constant, the ultraviolet cut-off and the renormalization mass scale, respectively. This inequality for $Z$ implies the positivity of the anomalous dimension in the one-loop approximation:
\begin{equation}
\gamma =\frac{1}{2}\mu\frac{\partial}{\partial\mu}\log{Z}=cg^2 \ge 0 .
\label{eqn:pos_anodim}
\end{equation}
The basic assumption of the unitarity bound is the positivity of the Hilbert space.
However, often this assumption does not hold in the covariant quantization of gauge theories, where the time component of the gauge field has a negative signature relative to the transverse components. The anomalous dimension of the electron in the one-loop approximation, for example, 
\begin{equation}
\gamma_e=\frac{e^2\alpha}{(4 \pi)^2}, \label{eqn:elanod}
\end{equation}
may be negative, depending on the choice of the gauge parameter $\alpha$. Although the total Hilbert space has an indefinite metric in gauge theories, the physical Hilbert space has a positive semidefinite norm and has a consistent probability interpretation \cite{Nishijima}. The anomalous dimensions of the physical operators, of course, satisfy positivity and are gauge independent.

We sometimes encounter negative anomalous dimensions in nonlinear sigma models (NL$\sigma$Ms), which are renormalizable in two dimensions \cite{HI, HI2}.
For example, the $O(N+1)$ invariant NL$\sigma$M, defined on the sphere $S^N$, is described by the Lagrangian
\begin{equation}
{\cal L}=\frac{1}{2}\cdot\frac{\left(\partial_{\mu}\vec{\varphi}\right)^2}{\left(1+\frac{g^2}{4}\vec{\varphi}^2\right)^2} \qquad
\left(\vec{\varphi}=(\varphi^1,\cdots, \varphi^N)\right),
\label{eqn:lagrangian1}
\end{equation}
and has the negative anomalous dimension
\begin{equation}
\gamma_{\varphi}=-\frac{N}{8\pi}g^2<0 .\label{eqn:ON1gamma}
\end{equation}
There are neither higher-order derivative terms, which often introduce a ghost with indefinite metric, nor gauge fields in this model. What is wrong in the proof of the unitarity bound of the wave function renormalization constant? This is the question we study in this report.

This paper is organized as follows.
In \S \ref{review}, we briefly review the unitarity bound of the wave function renormalization constant.
In \S \ref{nlsm}, we discuss the wave function renormalization constant in NL$\sigma$M in two dimensions.
In \S \ref{coordinate}, we describe the coordinate dependence of the anomalous dimension.
In \S \ref{summary}, we summarize our results.

%%%%%%%%%%%%%%%%%%%%%%%%%%%%%%%%%%%%%%%%%%%%%%%%%%%%%
\section{Unitarity bound of the wave function renormalization constant}\label{review}
%%%%%%%%%%%%%%%%%%%%%%%%%%%%%%%%%%%%%%%%%%%%%%%%%%%%%
Here we give a brief derivation of the unitarity bound of the wave function renormalization constant.
In the K\"{a}ll\'{e}n-Lehmann representation \cite{UK, KL, Nishijima, PS, Weinberg}, the commutation relation of the Heisenberg operators $\phi(x)$ and $\phi(y)$ is expressed as a superposition of the Jordan-Pauli function $i\Delta (x-y;m^2)$ for a particle with mass $m$:
\begin{eqnarray}
\bra{0} [\phi(x),\phi(y)] \ket{0}
=\int_0^\infty \rho(m^2) i\Delta (x-y;m^2) d m^2. \label{kallen-lehmann}
\end{eqnarray}
Here, the spectral function 
\begin{equation}
\rho (k^2)\equiv 2\pi \sum_n \delta ^{(2)}(p_n -k)|\bra{0} \phi(0)\ket{n}|^2\label{spectral_fn}
\end{equation}
depends only on the invariant mass squared, $m^2\equiv p_n^2$, of the complete set of states $\ket{n}$ and has support for non-negative $m^2$. The positivity of the metric of the Hilbert space implies the positivity of the spectral function:
\begin{equation}
\rho(m^2)\ge 0 .\label{positivity}
\end{equation}

The Jordan-Pauli invariant function $\Delta (x-y;m^2)$ is normalized according to 
\begin{equation}
\Delta(x,m^2)|_{x^0=0}=0,\qquad
\frac{\partial\Delta(x,m^2)}{\partial x^0} |_{x^0=0}=-\delta(x^1).\label{ETC}
\end{equation}
Now, the canonical commutation relations for the unrenormalized Heisenberg operators,
\begin{equation}
[\phi(t,x^1),\phi(t,y^1)]=0,\qquad 
[\phi(t,x^1),\dot{\phi}(t,y^1)]=i\delta(x^1-y^1),
\label{CCR}
\end{equation}
imply a sumrule for the spectral function. In fact, differentiating (\ref{kallen-lehmann}) with respect to $y^0$ and taking the equal-time limit, we obtain
\begin{equation}
\int_0^{\infty}\rho(m^2)dm^2=1.\label{sumrule}
\end{equation}

The spectral function $\rho(m^2)$ consists of an isolated delta function contribution from the one-particle state and a continuous spectrum contribution from scattering states of multiple particles:
\begin{equation}
\rho(m^2)=Z\delta(m^2-M^2)+\sigma(m^2).\label{sepation}
\end{equation}
Here, the wave function renormalization constant $Z$ represents the probability of finding a bare particle in the dressed particle state, while $\sigma(m^2)$ represents the continuous spectrum of multiple particles. The sum rule (\ref{sumrule}) for $\rho(m^2)$ reads
\begin{equation}
Z+\int \sigma(m^2)dm^2=1.\label{sumrule2}
\end{equation}

Since the positivity (\ref{positivity}) of $\rho(m^2)$ implies the positivity of both $Z\ge 0$ and $\sigma(m^2)\ge 0$, we find 
\begin{equation}
1-Z=\int \sigma(m^2)dm^2\ge 0 .\label{upperbound}
\end{equation}
The desired unitarity bound for the renormalization constant follows from 
this inequality and the positivity of $Z$:
\begin{equation}
0\leq Z \leq 1  .\label{unitarity}
\end{equation}
From this unitarity bound, we usually conclude that the anomalous dimension of a field $\phi$ is positive in the one-loop approximation, as shown in the introduction.
%%%%%%%%%%%%%%%%%%%%%%%%%%%%%%%%%%%%%%%%%%%%%%%%%%%%%%%%%%%%%%%%%%%%%%%%%
\section{Wave function renormalization constant in NL$\sigma$Ms in two dimensions}\label{nlsm}
%%%%%%%%%%%%%%%%%%%%%%%%%%%%%%%%%%%%%%%%%%%%%%%%%%%%%%%%%%%%%%%%%%%%%%%%%
Let us mow discuss the unitarity bound for the wave function renormalization constant in NL$\sigma$Ms described by the Lagrangian
\begin{equation}
{\cal L}=\frac{1}{2}g_{ij}(\varphi)\partial_{\mu}\varphi^i\partial^{\mu}\varphi^j.\label{nlsm_lagrangian}
\end{equation}
Here, we assume that the metric of the target space is positive definite to avoid the appearance of negative norm states. Furthermore, we do not incorporate higher-order derivative terms in our Lagrangian, since higher derivative terms inevitably introduce negative norm states. In the example discussed in the introduction, the target space is the coset space $S^N=SO(N+1)/SO(N)$, parameterized by $\vec{\varphi}=(\varphi^1,\cdots, \varphi^N)$, and the metric of the target space is given by
\begin{equation}
g_{ij}=\frac{\delta_{ij}}{\left(1+\frac{g^2}{4}\vec{\varphi}^2\right)^2}.
\qquad (i,j=1,2,\cdots,N)\label{metric}
\end{equation}
In general, a NL$\sigma$M is defined on a coset space $G/H$. It describes the interaction of massless Nambu-Goldstone bosons that appears when the symmetry group $G$ breaks down to its subgroup $H\subset G$. While the global symmetry $G$ is realized nonlinearly, its subgroup $H$ is realized linearly. We further assume that the fields $\vec{\varphi}=(\varphi^1,\cdots, \varphi^N)$ belong to some irreducible representation of the subgroup $H$.

With these assumptions, the K\"{a}ll\'{e}n-Lehmann representation can be generalized to \footnote{The vacuum is invariant under the unbroken symmetry group $H$.}
\begin{equation}
\bra{0} [\varphi^i(x),\varphi^j(y)] \ket{0}
=\delta^{ij}\int_0^\infty \rho(m^2) i\Delta (x-y;m^2) dm^2, \label{kallen-lehmann2}
\end{equation}
where the right-hand side is proportional to the $H$-invariant symmetric second-rank tensor $\delta^{ij}$, because the fields $\vec{\varphi}$ belong to an irreducible representation of $H$. If this were not the case, we would have several independent spectral functions for various representations.

The essential ingredient for the unitarity bound of the wave function renormalization constant is the positivity of the Hilbert space, as well as the canonical commutation relation (\ref{CCR}). In NL$\sigma$Ms, the derivative interaction term changes the definition of the canonical momentum, so that the equal-time commutation relation (\ref{CCR}) is no longer valid.
The canonical commutation relation in this case reads
\begin{equation}
[\varphi^i(t,x^1),g_{jk}\dot{\varphi}^k(t,y^1)]=i\delta^i_j\delta(x^1-y^1),\label{CCR2}
\end{equation}
or 
\begin{equation}
[\varphi^i(t,x^1),\dot{\varphi}^j(t,y^1)]=ig^{ij}\delta(x^1-y^1),\label{CCR3}
\end{equation}
where we have used the identity
\begin{equation}
g^{ik}g_{kj}=\delta^i_j.\label{raising}
\end{equation}
Taking the vacuum expectation value of (\ref{CCR3}), we find 
\begin{equation}
\bra{0}[\varphi^i(t,x^1),\dot{\varphi}^j(t,y^1)]\ket{0}
=i\bra{0}g^{ij}\ket{0}\delta(x^1-y^1).\label{VEVCCR}
\end{equation}
Because of the $H$-invariance of the vacuum, the right-hand side of this relation must be proportional to $\delta^{ij}$, and thus we have
\begin{equation}
\bra{0}g^{ij}(\varphi)\ket{0}=\delta^{ij}(1+B)\label{VEVmetric}
\end{equation}
for some quantity $B$. 
In the example of $S^N$, defined by Eq.(\ref{metric}), $B$ is given by 
\begin{equation}
B(S^N)=\bra{0}\left(1+\frac{g^2}{4}\vec{\varphi}^2\right)^2\ket{0}-1 \ge 0.\label{SNB}
\end{equation}

Differentiating (\ref{kallen-lehmann2}) with respect to $y^0$ and taking the equal-time limit, we obtain
\begin{equation}
\bra{0}[\varphi^i(t,x^1),\dot{\varphi}^j(t,y^1)]\ket{0}
=i\delta^{ij}\delta(x^1-y^1)\int\rho(m^2)dm^2.\label{VEVCCR2}
\end{equation}
Comparing this result with Eqs.(\ref{VEVCCR}) and (\ref{VEVmetric}), we find the sum-rule of the spectral function,
\begin{equation}
\int\rho(m^2)dm^2=1+B.\label{sumrule2}
\end{equation}
The expression (\ref{sepation}) of the spectral function as a sum of contributions from a single-particle and multi-particle states leads to 
\begin{eqnarray}
Z&=&1+B-C\leq 1+B, \label{sumrule_NLSM}\\
C&=&\int\sigma(m^2)dm^2 \ge 0.\label{multip}
\end{eqnarray}

We find from (\ref{sumrule_NLSM}) that when there are no derivative interactions, $B=0$ implies the unitarity bound $0\leq Z\leq 1$. In NL$\sigma$Ms, however, {\it the wave function renormalization constant $Z$ can be greater than $1$}, since $B$ may be positive, as shown in the example (\ref{SNB}).
Therefore, more generally, we have
\begin{equation}
0\leq Z \leq 1+B.\label{UB}
\end{equation}
The quantity $B$ comes from the matrix elements of local operators, and thus it represents the contributions of bubble diagrams, while $\sigma(m^2)$ gives the positive contributions of multi-particle states. 
\begin{figure}
\begin{center}
\resizebox{!}{30mm}{\includegraphics{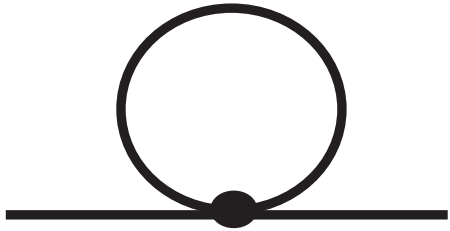}}
\end{center}
\caption{The lowest-order bubble diagram contributing to $B$}
\label{fig1}
\end{figure}
As an example, the lowest order contributions to $B$ and $C$ are shown in Figs.\ref{fig1} and \ref{fig2}. In the one-loop approximation, only $B$ in (\ref{SNB}) receives a nonvanishing contribution, \footnote{We use the Euclidean metric after the Wick rotation.}
\begin{equation}
B=\frac{g^2}{2}\int\frac{d^2k}{(2\pi)^2}\frac{N}{k^2}=\frac{Ng^2}{8\pi}\log{\frac{\Lambda^2}{\mu^2}},\label{SNB1}
\end{equation}
from the bubble diagram in Fig.\ref{fig1}, while $C$ vanishes in this approximation. This value of $B$ and Eqs.(\ref{eqn:pos_anodim}) and (\ref{sumrule_NLSM}) reproduce the negative one-loop anomalous dimension (\ref{eqn:ON1gamma}) given in the introduction. 
\begin{figure}
\begin{center}
\resizebox{!}{30mm}{\includegraphics{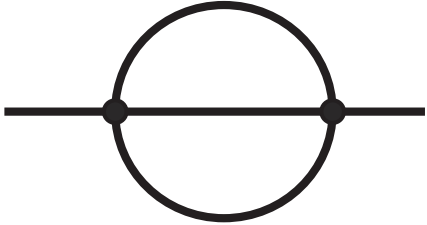}}
\end{center}
\caption{The lowest order contribution to $C$}
\label{fig2}
\end{figure}

It is worthwhile mentioning that the origin of the bound (\ref{UB}) is the invalid normalization (\ref{VEVCCR2}) of field variables. In fact, if we define a rescaled field $\tilde{\varphi}$ by
\begin{equation}
\varphi^i(x)\equiv \sqrt{1+B}\tilde{\varphi}^i(x),\label{rescaling}
\end{equation}
$\tilde{\varphi}(x)$ satisfies the equal-time commutation relation 
\begin{equation}
\bra{0}[\tilde{\varphi}^i(t,x^1),\dot{\tilde{\varphi}}^j(t,y^1)]\ket{0}
=i\delta^{ij}\delta(x^1-y^1).\label{NEWCCR}
\end{equation}
The spectral function $\tilde{\rho}(m^2)$ for the rescaled field $\tilde{\varphi}^i(x)$ satisfies the normal sumrule (\ref{sumrule}).
Therefore, we have the usual unitarity bound, 
\begin{equation}
0\le \tilde{Z}\le 1, \label{rescaledbound}
\end{equation}
for the rescaled wave function renormalization constant $\tilde{Z}$, defined by
\begin{equation}
\tilde{\rho}(m^2)=\tilde{Z}\delta(m^2-M^2)+\tilde{\sigma}(m^2).
\label{sepation2}
\end{equation}
The rescaling (\ref{rescaling}) has to be carried out order by order in perturbation theory, because $B$ depends on the coupling constant. 
The total renormalization constant $Z$ is proportional to $\tilde{Z}$,
\begin{equation}
Z=\tilde{Z}\cdot (1+B),\label{relation}
\end{equation}
and satisfies the bound (\ref{UB}).
%%%%%%%%%%%%%%%%%%%%%%%%%%%%%%%%%%%%%%%%%%%%%%%%%%%%%%%%%%%%
\section{Coordinate dependence of the anomalous dimension}\label{coordinate}
%%%%%%%%%%%%%%%%%%%%%%%%%%%%%%%%%%%%%%%%%%%%%%%%%%%%%%%%%%%%
The NL$\sigma$M is a field theory in which field variables take values on a curved manifold, like $S^N$. To describe the curved manifold, we have the freedom of choosing any coordinates we wish. In fact we used a stereographic projection to express $S^N$ in the coordinates of $N$-dimensional Euclidean space. Instead, we could use other parameterizations, and in this section, we use the simplest coordinates ($\phi^1, \phi^2, \cdots, \phi^{N+1}$) to express $S^N$, defined by
\begin{equation}
(\phi^1)^2+(\phi^2)^2+\cdots +(\phi^{N+1})^2=\frac{1}{g^2}.
\label{Nsphere}
\end{equation}
If we express $\phi^{N+1}$ in terms of the other variables, $\vec{\phi}=(\phi^1,\phi^2,\cdots,\phi^N)$, we have
\begin{equation}
\phi^{N+1}=\pm \sqrt{\frac{1}{g^2}-{\vec{\phi}}^2},
\label{phi_N+1}
\end{equation}
and the line element is given by
\begin{equation}
ds^2=\sum_{i=1}^{N+1}(d\phi^i)^2=\sum_{i,j=1}^Ng_{ij}d\phi^id\phi^j ,
\label{lineelement}
\end{equation}
where the metric in these coordinates reads
\begin{equation}
g_{ij}=\delta_{ij}+\frac{\phi^i\phi^j}{(\phi^{N+1})^2}.
\label{metric2}
\end{equation}
Because the inverse of this metric is
\begin{equation}
g^{ij}=\delta^{ij}-g^2\phi^i\phi^j,\label{invmetric}
\end{equation}
the one-loop contribution to $B$ defined by (\ref{VEVmetric}) is negative: 
\begin{equation}
B=-g^2\int\frac{d^2k}{(2\pi)^2}\frac{1}{k^2}=-\frac{g^2}{4\pi}\log\frac{\Lambda^2}{\mu^2}.\label{anotherb}
\end{equation}
This gives a positive anomalous dimension,
\begin{equation}
\gamma_{\phi}=\frac{g^2}{4\pi},
\label{anothergamma}
\end{equation}
in contrast to the negative anomalous dimension given in (\ref{eqn:ON1gamma}).

The sign of the anomalous dimension of the NL$\sigma$M depends on the parameterization of the manifold. How should we understand this fact? The anomalous dimension is an off-shell quantity describing the asymptotic behavior of the propagator in the ultraviolet region. Various parameterizations of a manifold give different metrics defining different field theories. There must be common quantities in these theories, since they all describe the same target manifold. In fact, under the coordinate transformation from $\vec{\varphi}$ with the metric (\ref{metric}) to $\vec{\phi}$ with the metric (\ref{metric2}), 
\begin{equation}
\vec{\phi}=\frac{\vec{\varphi}}{1+\frac{g^2}{4}{\vec{\varphi}}^2},\label{coordinatetr}
\end{equation}
physical quantities defined on the mass-shell are invariant. This is a special case of the theorem due to Kamefuchi, O'Raifeartaigh and Salam\cite{KOS}, which states that S-matrix elements are invariant under transformations of the type
\begin{equation}
\phi^i(x)=\varphi^i(x)+\sum_{jk} a^i_{jk}\varphi^j(x)\varphi^k(x)+
\sum_{jkl} a^i_{jkl}\varphi^j(x)\varphi^k(x)\varphi^l(x)+\cdots .
\label{kostr}
\end{equation}
S-matrix elements are defined as pole residues of Green's functions. Only the linear term in the transformation (\ref{kostr}) contributes to the pole residues. Quadratic and higher-order terms do not contribute to poles unless there are bound states. Therefore, all S-matrix elements, that is, all physical quantities, are invariant under the transformation (\ref{kostr}). 

In a previous work \cite{HI}, we derived the Wilsonian renormalization group equation in two-dimensional NL$\sigma$Ms with ${\cal N}=2$ supersymmetry. We used the so-called K\"{a}hler normal coordinate \cite{KNC} to derive the anomalous dimensions. Because the anomalous dimensions depend on the choice of the coordinate, we should interpret the results obtained in Ref\cite{HI} as the anomalous dimensions in the K\"{a}hler normal coordinate system.

%%%%%%%%%%%%%%%%%%%%%%%%%%%%%%%%%%%%%%%%%%%%%%%%%%%%%%
\section{Summary}\label{summary}
%%%%%%%%%%%%%%%%%%%%%%%%%%%%%%%%%%%%%%%%%%%%%%%%%%%%%%
The wave function renormalization constant $Z$ represents the probability of finding the bare particle in the physical particle, in the case that there are no derivative interactions. It satisfies the unitarity bound $0 \leq Z \leq 1$ in field theories without indefinite metric states.
This unitarity bound implies the positivity of the anomalous dimension of the field in the one-loop approximation. 

In NL$\sigma$Ms, however, this bound does not hold, due to the presence of the derivative interactions. The field dependence of the canonical momentum modifies the canonical commutation relation to
\begin{equation}
\bra{0}[\varphi^i(t,x^1),\dot{\varphi}^j(t,y^1)]\ket{0}
=i\delta^{ij}\delta(x^1-y^1)(1+B).\label{MODCCR}
\end{equation}
In this case, the actual unitarity bound of the wave function renormalization constant is 
\begin{equation}
0\leq Z \leq 1+B.\nonumber
\end{equation}
The contribution of the bubble diagrams to the self-energy diagrams, $B$, can be either positive or negative, while the contributions from multi-particle states, $C$, satisfies positivity, as expected. When $B$ is positive, the anomalous dimension in the one-loop approximation becomes negative. The sign of the contribution from the bubble diagrams, $B$, depends on the choice of the coordinates parameterizing the target manifold.

It is convenient to introduce the properly normalized field $\tilde{\varphi}$ by (\ref{rescaling}). The renormalization constant $\tilde{Z}$ of the field $\tilde{\varphi}$ satisfies the usual unitarity bound (\ref{rescaledbound}) and has a probabilistic interpretation.

%%%%%%%%%%%%%%%%%%%%%%%%%%%%%%%%%%%%%%%%%%%%%%%%%%%%%%
\section{Acknowledgements}
%%%%%%%%%%%%%%%%%%%%%%%%%%%%%%%%%%%%%%%%%%%%%%%%%%%%%%
This work was supported in part by Grants-in-Aid for Scientific
Research (\#13640283 and \#13135215) and a Sasakawa Scientific Research Grant from The Japan Science Society. We would like to thank Kazuhiko Nishijima, Thomas E. Clark and Muneto Nitta for useful communications. 

%%%%%%%%%%%%%%%%%%%%%%%%%%%%%%%%%%%%%%%%%%%%%%%%%%%%%%%%%%%%%%%%%%%%%%%%%%
%%%%%%%%%%%               References
%%%%%%%%%%%%%%%%%%%%%%%%%%%%%%%%%%%%%%%%%%%%%%%%%%%%%%%%%%%%%%%%%%%%%%%%%%


\begin{thebibliography}{99}

\bibitem{UK} 
H.~Umezawa and S.~Kamefuchi, Prog. Theor. Phys. {\bf 6} (1951) 543.

\bibitem{KL}
G.K\"{a}ll\'{e}n, Helv. Phys. Acta, {\bf 25} (1952) 417\\
H.~Lehmann, Nuovo Cimento {\bf 11} (1954) 342.

\bibitem{Nishijima} K.~Nishijima, Fields and Particles (W.A.~Benjamin Inc., New York, 1968).

\bibitem{PS} M.~Peskin and D.~Schroeder, An Introduction to Quantum Field Theory (Westview Press, Boulder, 1995).

\bibitem{Weinberg} S.~Weinberg, The Quantum Theory of Fields (Cambridge University Press, Cambridge, 1995).

\bibitem{HI} K.~Higashijima and E.~Itou, 
Prog. Theor. Phys.  {\bf 108} (2002) 737,
 hep-th/0205036
 
\bibitem{HI2} K.~Higashijima and E.~Itou, 
Prog. Theor. Phys.  {\bf 109} (2003) 751,
 hep-th/0302090.

\bibitem{KOS} S.~Kamefuchi, L.~O'Raifeartaigh and A.~Salam, Nucl. Phys. {\bf 28} (1961) 529.

\bibitem{KNC}
K.~Higashijima and M.~Nitta, Prog. Theor. Phys. {\bf 105} (2001) 243, hep-th/0006027. \\
K.~Higashijima, E.~Itou and M.~Nitta, Prog. Theor. Phys. {\bf 108} (2002) 185, hep-th/0203081.

\end{thebibliography}
\end{document}